\documentclass[12pt]{iopart}
\pdfoutput=1

\usepackage{url}
\usepackage{graphicx}
\usepackage{color}
\usepackage{latexsym}
\usepackage{enumerate}
\usepackage{multirow}
\usepackage{twoopt}
\usepackage{array}
\usepackage{alltt}

\definecolor{orange}{rgb}{0.8,0.2,0.2}

\newcommand{\barT}{T}
\newcommand{\WTA}{\textit{WTA}}

\begin{document}

\title[Quantum-Classical Transitions in Complex Networks]{Quantum-Classical Transitions in Complex Networks}

\author{Marco Alberto Javarone}

\address{DIEE - Dept. of Electrical and Electronic Engineering\\ University of Cagliari, P.zza D'armi 09123
Cagliari, Italy}
\ead{marcojavarone@gmail.com}

\author{Giuliano Armano}

\address{DIEE - Dept. of Electrical and Electronic Engineering\\ University of Cagliari, P.zza D'armi 09123
Cagliari, Italy}
\ead{armano@diee.unica.it}

\begin{abstract}
The inherent properties of specific physical systems can be used as metaphors for investigation of the behavior of complex networks. This insight has already been put into practice in previous work, e.g., studying the network evolution in terms of phase transitions of quantum gases or representing distances among nodes as if they were particle energies. This paper shows that the emergence of different structures in complex networks, such as the scale-free and the winner-takes-all networks, can be represented in terms of a quantum–classical transition for quantum gases. In particular, we propose a model of fermionic networks that allows us to investigate the network evolution and its dependence on the system temperature. Simulations, performed in accordance with the cited model, clearly highlight the separation between classical random and winner- takes-all networks, in full correspondence with the separation between classical and quantum regions for quantum gases. We deem this model useful for the analysis of synthetic and real complex networks.
\end{abstract}

\maketitle

\section{Introduction}
Many natural and man-made complex systems can be modeled by complex networks, i.e., biological cells and their interactions, social dynamics, the World Wide Web, and any other system containing a notable number of interacting elements~\cite{albert02,guimer03}.
Notably, complex networks allow one to represent systems at a non-equilibrium state, meaning that new nodes and/or new links can be added over time during an evolutionary, often irreversible, process. Albert and Barabasi~\cite{albert02} illustrated the central role of statistical mechanics for the analysis of these networks. One of the first models for random graph dynamics was developed by Erdos and Renyi, see~\cite{erdos01}. This kind of network, called an \textit{E–R graph}, has a low clustering coefficient and a binomial degree distribution of nodes, converging to a Poissonian distribution for a large number of nodes. In their \textit{WS model}, Watts and Strogatz~\cite{watts01} interpolated \textit{E-R graphs} with a regular ring lattice, achieving networks characterized by short average path lengths, high clustering coefficient, and homogeneous degree distribution of nodes.
Observing that in many real networks the degree distribution of nodes follows a power-law, Barabasi and Albert developed the \textit{BA model} \cite{albert02}, defining the concept of \textit{scale-free} networks.
The power-law equation is
\begin{equation} \label{eq:plaw}
P(k) \sim c\cdot k^{-\gamma}
\end{equation}
\noindent where $k$ is the node degree, $c$ is a normalizing constant and $\gamma$ is a parameter of the distribution known as the scaling parameter. The scaling parameter usually lies in the range $\left [ 2,3 \right ]$. A notable consequence is that, in these networks, only few nodes (called hubs) have many links. 
The evolutionary process of networks has been studied by many authors, often resorting to theoretical physics ---in particular, mapping network evolution to quantum mechanics.
For the sake of brevity, let us cite only a few.
Bianconi and Barabasi \cite{bianconi01} compared Bose-Einstein Condensation phenomena (BEC) to \textit{winner-takes-all} policies (\WTA~hereinafter); Bianconi \cite{bianconi02} discussed the symmetric construction of bosonic and fermionic networks, asserting that the former networks are scale-free and the latter are growing Cayley trees.
Kriukov et al. \cite{krioukov01} developed a geometric framework to study the structure and functions of complex networks, interpreting edges as non-interacting fermions whose energies are hyperbolic distances between nodes.
Shen Yi et al. \cite{shen01} discussed an inverse approach to network evolution defining a relation with their model and Fermi-Dirac statistics.
Baronchelli et al. \cite{baronchelli01} defined a framework using bosonic reaction-diffusion processes, with the aim of analysing dynamical systems on complex networks.
Perseguers et al. \cite{perseguers01} developed a model of quantum complex networks, drawing a link between the field of complex networks and that of quantum computing.

In this paper we propose a theoretical model of network evolution inspired by the physics of fermions, obtained by mapping complex networks to fermionic gases.
The proposed model, which can be seen as dual with respect to bosonic networks (\cite{bianconi01}), 
shows that the emergence of different network structures can be represented in terms of a quantum–classical transition for quantum gases. Furthermore, we study the evolution of complex networks composed of elements characterized by their own fitness parameters; these are inherent competitive factors embedded in each node.
The remainder of the paper is organized as follows: Section~\ref{sec:quantum-stat} gives a brief introduction to quantum statistics. Section~\ref{sec:networks-models} describes some existing models of networks as non-interacting particle systems. Section~\ref{sec:network-fds} introduces the proposed model of fermionic networks. Section~\ref{sec:results} shows the results of the corresponding simulations. Conclusions (Section~\ref{sec:conclusions}) end the paper. 
\section{Quantum Statistics} \label{sec:quantum-stat}
Statistical mechanics assumes a central role when dealing with systems composed of many particles, the underlying assumption being that the particles are identical and indistinguishable. Moreover, their quantum energy levels are extremely closely spaced, with a cardinality much greater then the number of particles. Energy levels can be grouped in bundles with the approximation that levels in the same bundle have the same energy. Particles with a symmetric wavefunction, called bosons, obey Bose-Einstein statistics, whereas particles with an antisymmetric wavefunction, called fermions, obey Fermi-Dirac statistics~\cite{huang01}.
Given a system with $N$ particles of the same type, we can build an N-body wavefunction, with several admissible states. For each state $\alpha$, the corresponding number of particles, say $n_{\alpha}$ (also called the occupation number), is given by the following equation:
\begin{equation} \label{eq:occnumber}
n_{\alpha} = \cases{0,1,...,\infty & \mbox{bosons } \\ 
                                          0,1 & \mbox{fermions}}
\end{equation}
and $\sum_{\alpha} n_{\alpha} = N$. Considering a gas composed by $N$ bosons, the number of microstates is computable according to the equation
\begin{equation} \label{eq:bosons_arr}
\Omega_{b} = \Pi_{i} \frac{(n_{i} + g_{i})!}{n_{i}!\,g_{i}!}
\end{equation}
with $g_{i}$ representing the $i$th bundle. The distribution of particles follows the Bose-Einstein statistics
\begin{equation} \label{eq:bosons}
n_{i}^{b} = g_{i} \cdot {\left ( e^\frac{{\epsilon_{i} - \mu}}{k_{b}T} - 1 \right ) }^{-1}
\end{equation}

\noindent where $\epsilon_{i}$ denotes the energy of the $i$th bundle, $\mu$ the chemical potential, and $k_{b}$ the Boltzmann constant.
In the event that a gas is composed of fermions, we must consider also the Pauli exclusion principle. Hence, the number of microstates is computable according to the equation
\begin{equation} \label{eq:fermions_arr}
\Omega_{f} = \Pi_{i} \frac{g_{i}!}{n_{i}! \, (g_{i} - n_{i})!}
\end{equation}
Here, the distribution of particles follows the Fermi-Dirac statistics,
\begin{equation} \label{eq:fermions}
n_{i}^{f} = g_{i} \cdot \left ( e^\frac{{\epsilon_{i} - \mu}}{k_{b}T} + 1 \right )^{-1}
\end{equation}
Both these distributions approximate the classical behaviour in proximity of the high-temperature limit, showing a quantum-classical transition. This phenomenon occurs when particles sparsely occupy excited states. In particular, with $\lambda$ thermal wavelength and $\rho$ density, the following conditions hold:

\begin{equation}\label{eq:transition_qc}
\cases{\rho \lambda^{3} \gg 1  & \mbox{classical regime} \\
\rho \lambda^{3} \approx 1  & \mbox{onset of quantum effects}}
\end{equation}

The classical regime is described by the Maxwell-Boltzmann distribution. In particular, with $Z =  \sum_{j} g_{j}e^{-\frac{\epsilon_{j}}{k_{b}T}}$ partition function, we write
\begin{equation} \label{eq:maxwell}
n_{i}^{mb} = \frac{N}{Z} \cdot g_{i} \cdot e^{-\frac{\epsilon_{i}}{k_{b}T}}
\end{equation}

\section{Networks as Particle Systems} \label{sec:networks-models}
Under certain assumptions, complex networks can be considered as thermodynamic systems that evolve from one state to another~\cite{hartonen01}. In this section we give a synthetic overview of some existing models of network dynamics based on quantum statistics.

\subsection{Bosonic Networks} 
In this model, Bianconi and Barabasi \cite{bianconi01} compared network evolution to a phase transition of bosonic gases. Two main structures, i.e., \textit{fit-get-rich} and \WTA, are identified as two different phases at low temperature. In this model, each node is interpreted as an energy level and each link as a pair of particles. 
Starting from a fitness parameter $\eta$, energy is computed according to the following equation:
\begin{equation} \label{eq:energy_fitness}
\epsilon = -\frac{1}{\beta} \cdot \log \eta
\end{equation}
with $\beta = \frac{1}{T}$. Here, the fitness parameter $\eta$ describes the ability of each node to compete for new links. In particular, for the $i$th node, the probability of connection with new nodes is proportional to:
\begin{equation} \label{eq:prob_link}
\Pi_{i} = \frac{\eta_{i} k_{i}}{\sum_{j} \eta_{j}k_{j}}
\end{equation}
with $k_{i}$ degree of the $i$th node. Notably, new nodes tend to link with pre-existing nodes having high values of $(\eta,k)$. 
The generation of a scale-free network in the \textit{fit-get-rich} phase is characterized by Eq.~(\ref{eq:plaw}) and entails the presence of a few nodes with a high degree connected to many others with low degree. 
In a bosonic gas, when the temperature decreases, particles aim to occupy lower energy levels. Then, at a temperature below the critical temperature $T_{c}$, Bose-Einstein condensation takes place. In this model, as the temperature decreases, many particles move to lower levels while keeping the corresponding particles at upper levels. In so doing, links concentrate on a few nodes, until they condensate in the \WTA~phase, characterized by the fact that only one node predominates.
In \cite{bianconi02}, Bianconi discussed the differences between bosonic and fermionic networks, showing that the former are scale-free, whereas the latter can be represented by Cayley trees.

\subsection{Fermi-Dirac Statistics of Complex Networks}
In~\cite{shen01}, the authors proposed an inverse approach to network evolution, starting from a random network with an average degree equal to $2m$. Using an \textit{illness model}, at each step, one randomly-selected node becomes ``ill'', then its illness causes a loss of links. An illness parameter $I$ is defined and assigned randomly to each node. The probability for each node to lose a link depends on $I$ and on the node degree. Hence, the number of links decreases with time. Each node has an energy defined by
\begin{equation} \label{eq:energy_ill}
\epsilon = -\frac{1}{\beta} \log I
\end{equation}
\noindent with $\beta = \frac{1}{k_{b}T}$. A link between two nodes corresponds to a couple of non-interacting particles placed in two energy levels whose value is computed by Eq.~\ref{eq:energy_ill}.
When a new node joins the network, a new energy level is added, with $m$ new particles, and other $m$ links are deleted, due to the illness. The authors showed that the behavior of such system can be approximated by a Fermi distribution, so that the network can be seen as a Fermi gas.

\section{Fermionic Networks}\label{sec:network-fds}
Let us now introduce a novel proposal for modeling network dynamics, inspired from the physics of fermions. 
Given a network $G=(V,E)$, with $V$ nonempty set of nodes and $E$ nonempty set of links, let us represent each link as a particle and each node as a degenerate bundle of energy levels.
Usually, the number $g_{i}$ of available states in the $i$th bundle is much larger than the number $p_{i}$ of its particles.
Let us assume that the $i$th bundle has an energy $\epsilon_{i}$. This value can be assigned randomly or depend on a property of the system ---e.g. a fitness parameter $\eta$ or any other function deemed relevant, with the trivial constraint that it must be computable for each node of the network.
In the proposed model, lower bundles have more energy levels. In particular, the first bundle has $n-1$ levels, the second has $n-2$ levels, and so on.
Note that the link $l_{ij}$, which connects nodes $i$ and $j$, is represented only by a single energy level, i.e., $\epsilon_{ij}$, which in turn belongs to the $i$th bundle (under the assumption that the $i$th bundle is deeper than the $j$th one).
In so doing, the last node, say $y_0$, is represented by a bundle without energy levels, although it can be linked in the event that a particle stays at the $\epsilon_{xy_0}$ level, with $x$ corresponding to one of the other nodes. 
\subsection{Modeling Network Evolution} 
\begin{figure}[!ht]
\centering
\includegraphics[width=2.5in]{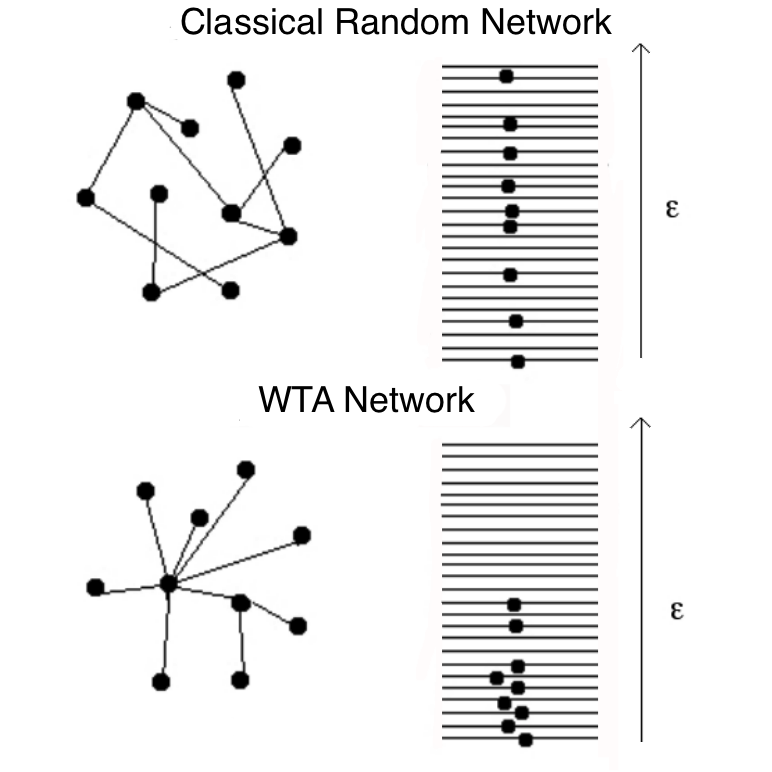}
\caption{\label{fig:fermionic_network_scale} Left, from top to bottom, the evolution of a network with $10$ nodes and $9$ links from a classical random network to a \WTA~network. Right: their corresponding fermionic models, which result from a cooling process that pushes particles to low-energy levels.}
\end{figure}

Let us consider an evolving network, i.e., a network that changes over time. Almost all real networks evolve over time; examples are social networks (where people find or lose friends or co-workers) and the web (where web-sites compete to gain more inlinks). Furthermore, let us consider this network as a closed system, so that the number of nodes and the number of links remain constant over time.
As discussed before, for every node, a bundle is defined ---whose energy is computed with Eq.~(\ref{eq:energy_fitness}). In so doing, the relative position of each bundle depends on the value of its energy, so that deeper bundles embody more states. 
Considering the ability of the particles to jump between energy levels as the temperature varies, at high temperatures particles follow the classical Maxwell-Boltzmann distribution, being spread among the available states according to  Eq.~(\ref{eq:maxwell}).
On the other hand, as the temperature decreases, many particles move to lower energy levels (see Figure~\ref{fig:fermionic_network_scale}).

In this work, we consider the evolution of a system caused by cooling and heating processes. A detailed analysis of both processes follows.

\subsubsection{Cooling Process}
During a cooling process, a few nodes gain new links and their degree $k_{i}$ is increased. Given the number of particles, it is possible to compute the Fermi Energy $E_{f}$ of the system as the energy of the bundle containing the last particle at $T \to 0$.
Hence, as the temperature decreases, and assuming that the number of particles approximates the number of bundles, the \WTA~phase takes place (see also \cite{bianconi01}). 
For every variation of the temperature, the probability for a particle to jump from the $i$th to the $j$th bundle is computed according to the following equation:
\begin{equation} \label{eq:jump_probability}
p(i \to j) = \frac{\Delta T}{\barT} \cdot \frac{1}{\Delta B(j,i)} \cdot  f(g_{j})
\end{equation}
\noindent where $\barT$ denotes the temperature of the system before the variation, $\Delta T$ is the variation of the temperature, $\Delta B(j,i)$ is the distance between bundles $j$ and $i$, and $f(g_{j})$ is the function
\begin{equation} \label{eq:free_space}
f(g_{j}) = \cases{0 & \mbox{if  $g_{j} = 0$} \\ 
                                  1 & \mbox{if  $g_{j} \ge 1$}}
\end{equation}
with $g_{j}$ number of available states in the $j$-th bundle.
Hence, considering that a particle in the $i$th bundle can jump to $i - 1$ underlying bundles, each with a probability given by Eq.~(\ref{eq:jump_probability}), the probability $p_J$ to jump from the $i$th to another bundle is
\begin{equation} \label{eq:total_jump_probability}
p_{J}(i) = \sum_{z=1}^{i-1} p(i \to z)
\end{equation}
and the probability $p_S$ to stay in the same bundle is 
\begin{equation} \label{eq:stay_probability}
p_{S}(i) = 1 - p_{J}(i)
\end{equation}
Then, the final bundle of each particle is chosen by a weighted random selection among all candidate bundles (including the bundle in which the particle is located). 
\subsubsection{Heating Process}
Heating is performed after cooling. Particles can now move to higher energy levels, gradually generating vacancies at lower levels. Also in this case, for every variation of the temperature, the probability for a particle in the $i$th bundle to jump to the $j$th bundle is computed using a variant of Eq.~(\ref{eq:jump_probability}), in which $f(g_{j})$ is defined as

\begin{equation} \label{eq:free_space_heat}
f(g_{j}) = \cases{0 & \mbox{if  $g_{j} = 0$} \\ 
                                  1 - \frac{p_{j}}{g_{j}} & \mbox{if  $g_{j} \ge 1$}}
\end{equation}
\noindent with $p_j$ number of particles located at the $j$th bundle. Eq.~(\ref{eq:free_space_heat}) has been devised to prevent that, at high temperatures, particles from filling few high-energy levels densely.
For each particle, the probability of jumping is computed by the following equation:
\begin{equation} \label{eq:total_jump_probability_heat}
p_{J}(i) = \sum_{z=i + 1}^{n-1} p(i \to z)
\end{equation}
and the probability of staying by Eq.~(\ref{eq:stay_probability}).
The same criterion adopted in the cooling process (i.e., weighted random selection) is applied to choose the energy level of each particle.
In our model, the temperature corresponds to a phenomenon that leads to an evolution (e.g., the evolution of relations among people or a competition for new inlinks among web-sites). 
To complete the model, let us assume that each network has the structure of an E-R graph when generated at time $t=0$. 

\section{Results}\label{sec:results}
\begin{figure}[!ht]
\centering
\includegraphics[width=5.8in]{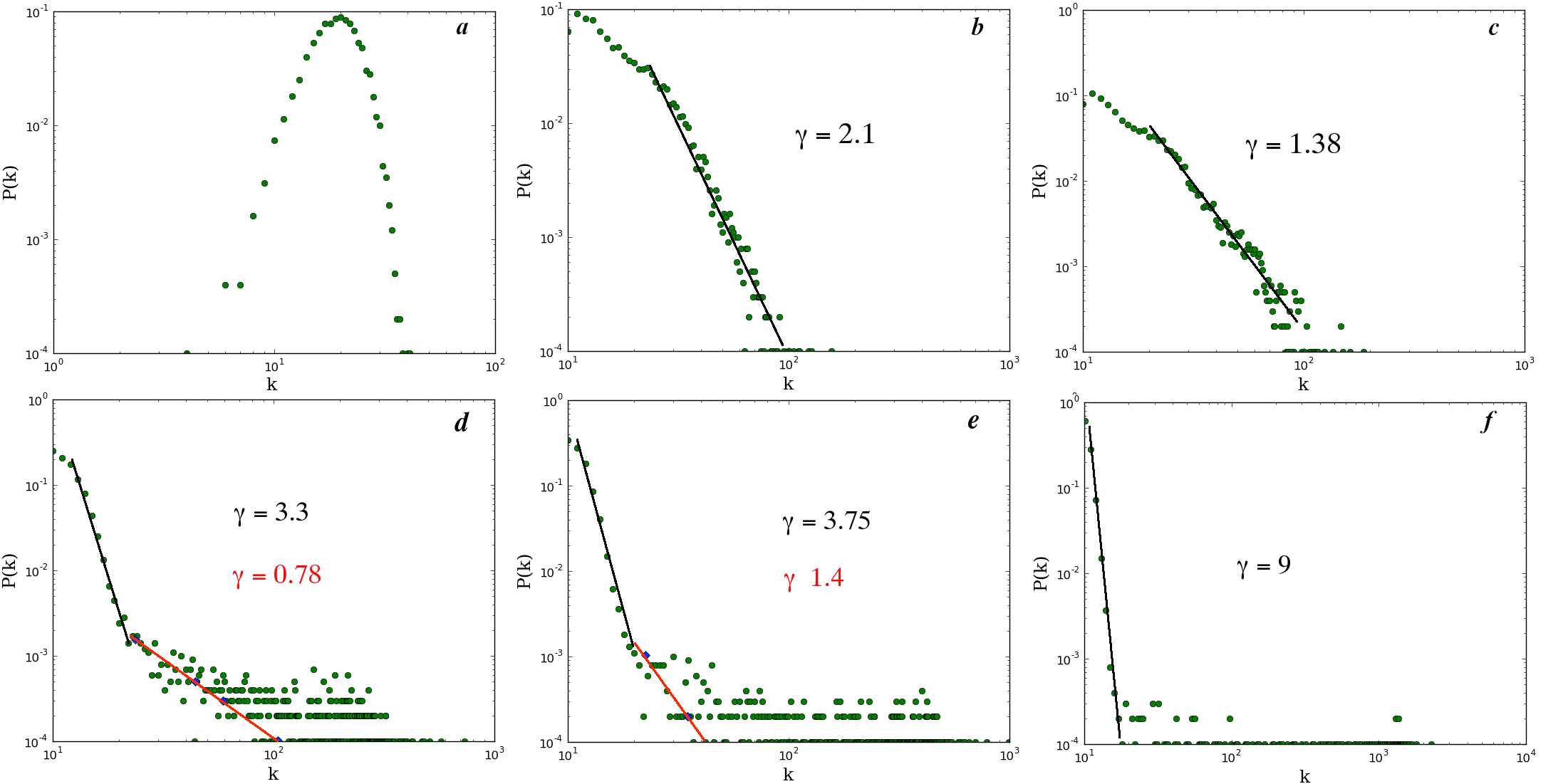}
\caption{\label{fig:er2sf} The evolution of the degree distribution of a network, during a cooling process, with $10000$ nodes and $\langle k \rangle = 20$, generated at $T = 100$K. Each panel shows the network at different time steps $t$: \textbf{a)} at $t=0$; \textbf{b)} at  $t=4$; \textbf{c)} at $t=5$; \textbf{d)} at $t=19$; \textbf{e)} at $t=28$; \textbf{f)} at $t=50$. Note that for $t=0$ the network has an \textit{E-R graph} structure, whereas for $t=50$ it has a \WTA~structure. Continuous black and red lines are used to highlight data interpolation. The corresponding scaling parameter(s) $\gamma$ is (are) indicated in each panel.}
\end{figure}

\begin{figure}[!ht]
\centering
\includegraphics[width=5.8in]{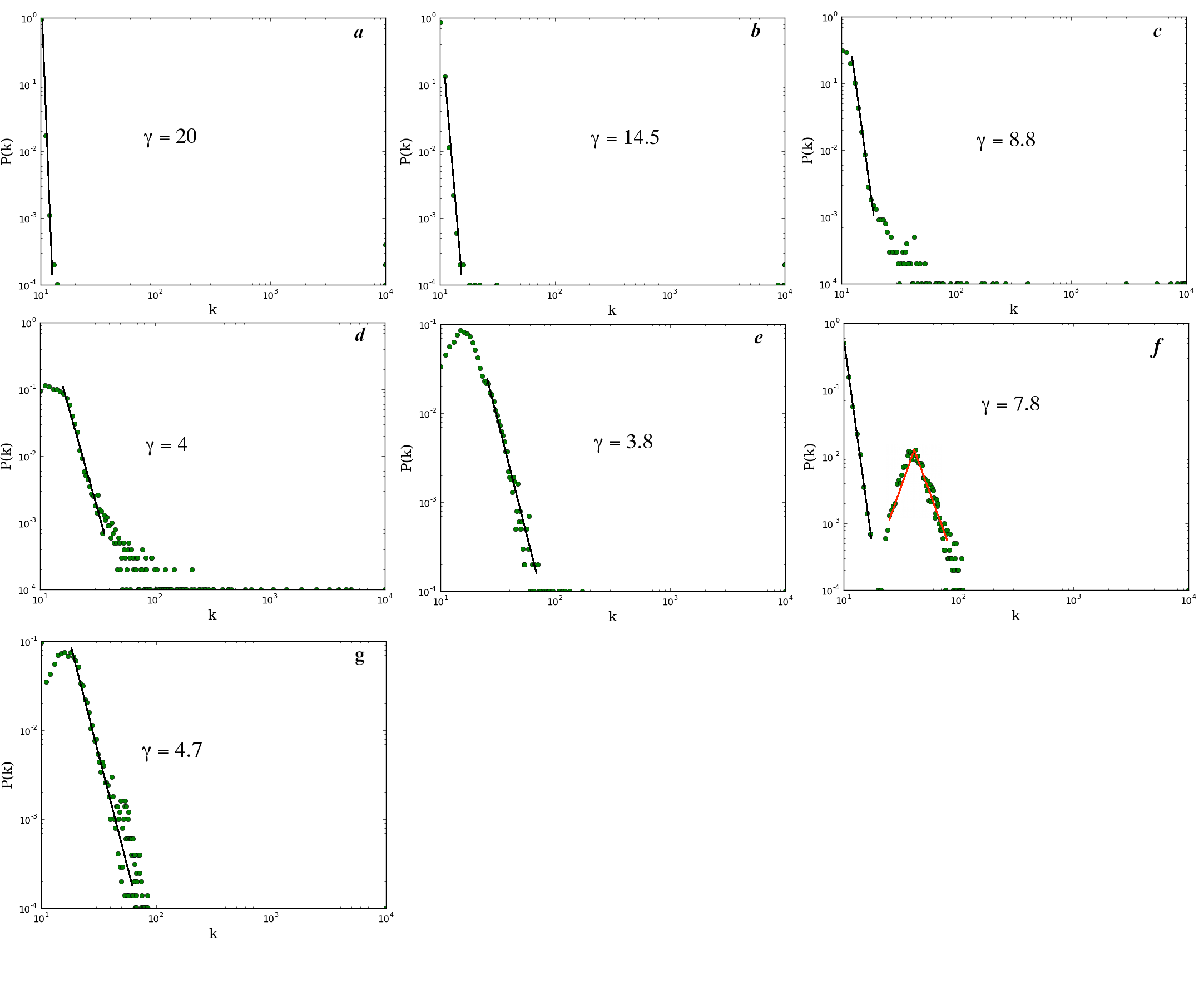}
\caption{\label{fig:network_heating} The evolution of the degree distribution of a network, during a heating process, with $10000$ nodes and $\langle k \rangle = 20$. Each panel shows the network at a different time step $t$: \textbf{a)} at $t=0$; \textbf{b)} at $t=15$; \textbf{c)} at $t=28$; \textbf{d)} at $t=34$; \textbf{e)} at $t=40$; \textbf{f)} at $t=58$; \textbf{g)} at $t=65$. Note that for $t=0$ the network has a \WTA~structure. Continuous black and red lines are used to highlight data interpolation. The corresponding scaling parameter $\gamma$ is indicated in each panel.}
\end{figure}

\begin{figure}[!ht]
\centering
\includegraphics[width=4in]{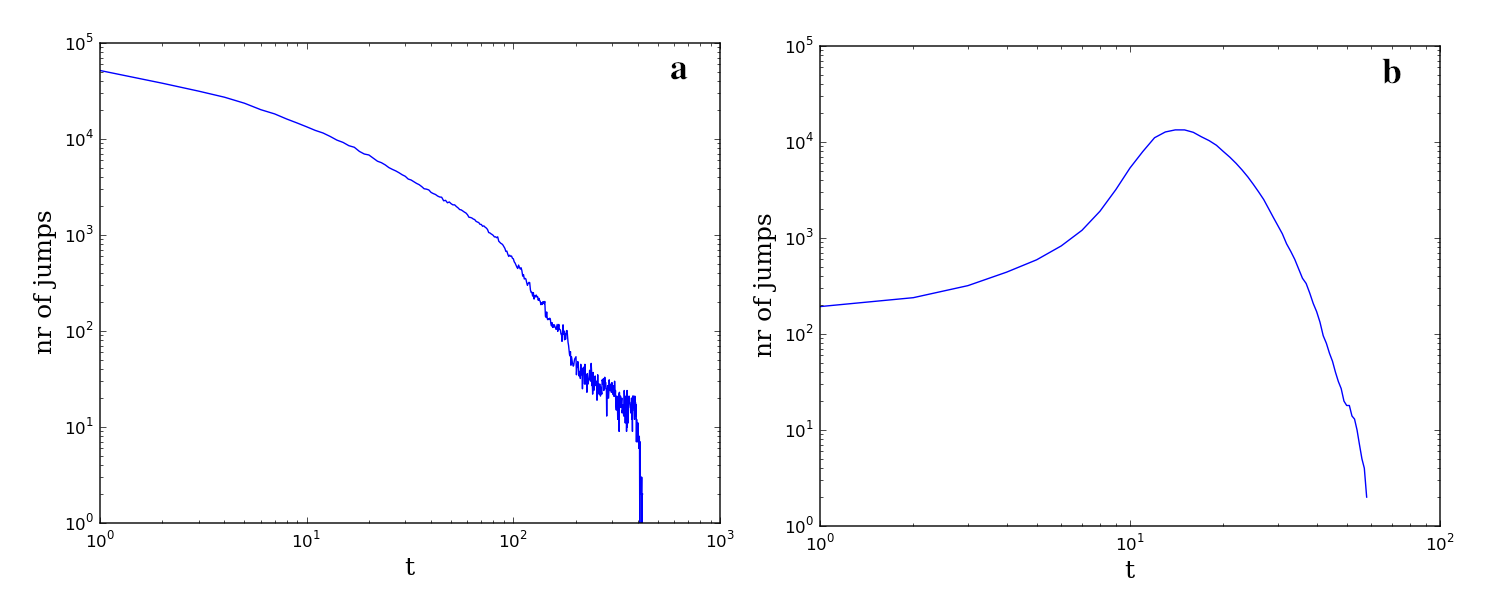}
\caption{\label{fig:network_jump} The number of particles that change their energy level (indicated as number of jumps) with time (considering a network with $n = 3000$). \textbf{a} During the cooling process, the number of jumps rapidly decreases. \textbf{b} During the heating process, at the beginning, all particles are constrained to low-energy levels. 
After a few time steps particles can find more available states in the upper bundles and the number of jumps increases. 
This curve reaches its maximum when all particles have available upper energy levels to reach, until these top levels become full and the number of jumps begins to decrease. At the end, all the particles are mainly arranged in the higher energy levels.}
\end{figure}

The proposed fermionic model has been tested with many simulations. In particular, we generated networks of different sizes with an E-R graph structure. These networks have been implemented by connecting nodes randomly ---giving rise to a graph $G(n,\zeta)$, where $n$ is the number of nodes and $\zeta$ is the probability of an edge to be drawn (note that an edge is drawn independently of other edges). 
Their degree distribution is binomial, converging to a Poissonian distribution for a large number of nodes, according to the following definition:
\begin{equation} \label{eq:poisson}
P(k) \sim e^{-\zeta n} \cdot \frac{(\zeta n)^{k}}{k!}
\end{equation} 
Simulations have been performed with a number of nodes ranging from $50$ to $10000$, $\zeta = \frac{\langle k \rangle}{n - 1}$ with $\langle k \rangle$ average degree of the network (see \cite{newman01}) and an initial temperature ranging from $100$ to $500$K. 
For each simulation the network evolves until all particles of the model reach their final position, for both the cooling and heating process.
At each time step, the temperature is increased (heating) or decreased (cooling) by $10 \%$, then the algorithm computes the new positions of the particles and analyzes the degree distribution, computing the scaling parameter $\gamma$ and the normalizing costant $c$. Scaling parameters have been estimated, as suggested in~\cite{clauset01}, by using the following equation:
\begin{equation}
\hat{\gamma} = 1 + n \cdot \left [ \sum_{i=1}^{n} \ln \frac{k_{i}}{k_{m}} \right ]^{-1}
\end{equation}
\noindent with $k_{m}$ minimum degree estimated. The normalizing constant is computed as follows:
\begin{equation}
c = \frac{1}{\int \limits_{k_{m}}^\infty k^{-\gamma} dk}
\end{equation}
Figure~\ref{fig:er2sf} illustrates a transition between the \textit{E-R graph} structure and the \WTA~structure, for a network with $10000$ nodes and $\langle k \rangle = 20$, generated at $100$K. 
As shown in this figure, a cooling process in an \textit{E-R} graph entails a transition to a scale-free structure after four time steps. After $19$ time steps all networks apparently converge to a \WTA~structure, showing in some cases composite distributions, which in turn can be identified by a process of logarithm data binning (see~\cite{milojevic01}).
\begin{figure}[!ht]
\centering
\includegraphics[width=3in]{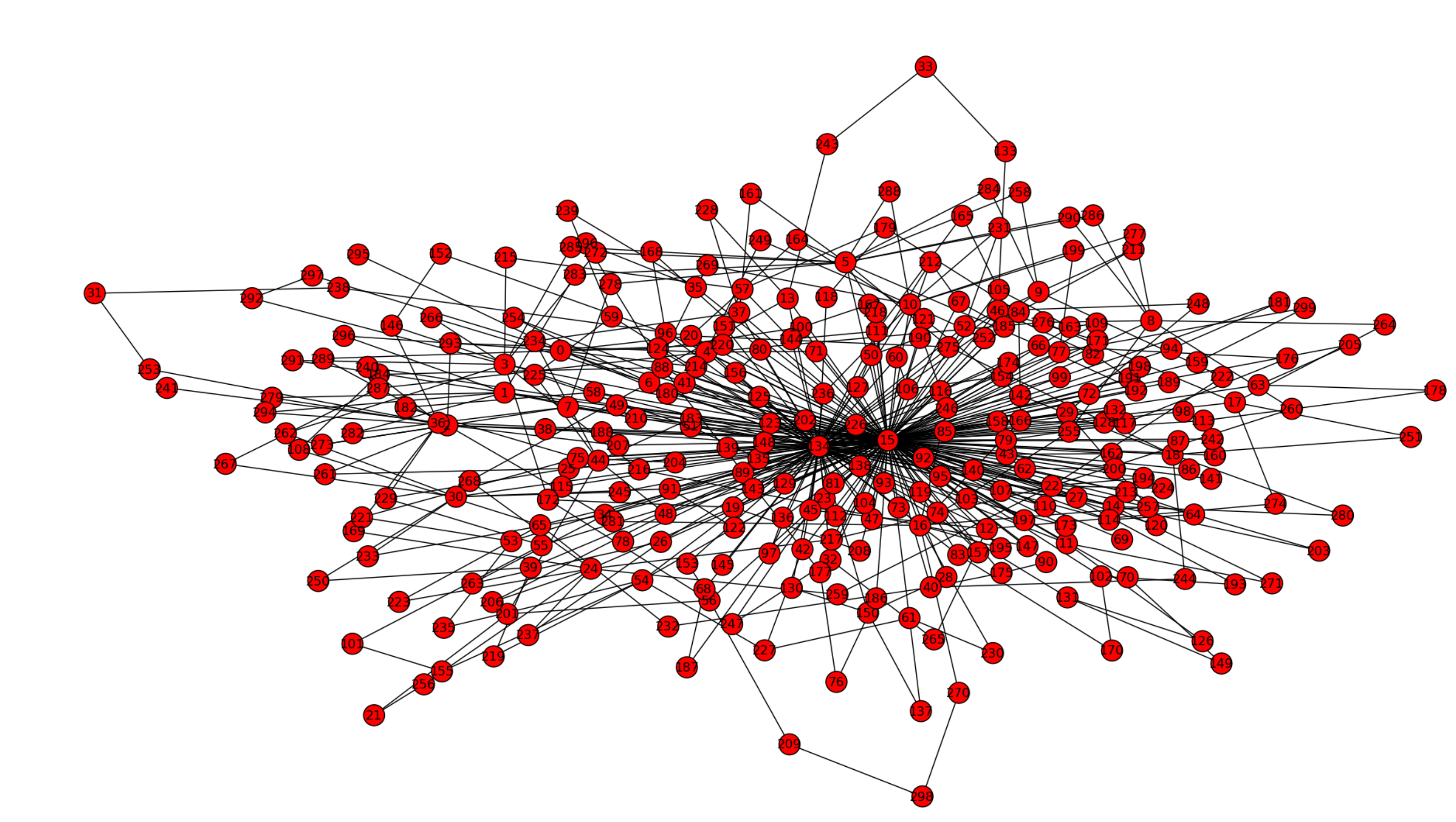}
\caption{\label{fig:network_wta} A network with $n=300$ with a \WTA~structure, obtained by cooling an \textit{E-R} graph until $T \approx 0$K. As highlighted by the figure, there are a few winning nodes (clearly visible in the center of the figure).}
\end{figure}
A small network having a \WTA~structure is shown in Figure~\ref{fig:network_wta}, where only a few nodes have a great amount of links, i.e., their bundles contain the majority of particles.
After the cooling process, the network with $10000$ nodes is subject to the heating process and Figure~\ref{fig:network_heating} illustrates the evolution of the degree distribution.
During the heating process (see Figure~\ref{fig:network_heating}), the network loses its \WTA~structure. Then, its degree distribution apparently becomes scale-free (see panel \textbf{d} and \textbf{e} of Figure~\ref{fig:network_heating}). As the temperature increases further, it converges to a hybrid distribution (see panel \textbf{f} of Figure~\ref{fig:network_heating}) characterized by two main distributions: exponential and Gaussian.
Eventually, a homogeneous structure ($\gamma = 4.7$) emerges. Similar results have been achieved in all simulations, varying the number of nodes and considering different initial temperatures. 
Figure~\ref{fig:network_jump} shows the number of particles that, at each time step during both processes, change their energy level.
As the temperature decreases, particles move to lower energy levels until they occupy the deeper bundles; then the number of particles that change their position falls to zero. On the other hand, while heating the system, particles slowly begin to jump to higher energy levels. At the beginning of this process, only a few particles move, as the majority of particles are in fact constrained to their level due to the lack of available (close) upper levels. Then, all particles can move and the number of jumps reaches a maximum, until the upper levels also begin to fill. At the end of the process, all particles mainly occupy the upper energy levels and the number of jumps falls to zero.
\subsection{Discussion}
Fermionic networks show that the emergence of different structures can be represented as a quantum-classical transition for quantum gases.
In particular, a \WTA~structure corresponds to a fermionic gas approximated by the quantum regime at low temperatures.
On the other hand, a classical random network corresponds to the same gas in the classical regime at high temperatures. During a cooling process, at intermediate temperatures, a scale-free structure emerges. As shown in Figure~\ref{fig:er2sf}, the \textit{E-R} structure rapidly changes into a scale-free, with a scaling parameter of about $2.1$. 
This parameter before decreases upto $1.38$ and afterwards increases until the network loses its neat scale-free structure (see panel \textbf{d} of Figure~\ref{fig:er2sf}) and begins to converge to the \WTA~structure characterized by a high value of the scaling parameter.
In particular, a homogeneous structure emerges, with the presence of hubs (i.e., nodes with high degree). At the end of the cooling process, a few nodes have a very high degree ($\sim n$) and the remaining nodes have low degree.
Considering the heating process, we observed that the scaling parameter slowly decreases at the beginning of the process. 
During the first simulation steps the network apparently converges to a scale-free structure, while for values of the scaling parameter of around $3.8 - 3.5$ the network converges to a hybrid structure, which follows an exponential distribution for low values of $k$ and a Gaussian distribution for high values of $k$ (see panel \textbf{f} of Figure~\ref{fig:network_heating}). Eventually, a homogeneous structure takes over at high temperatures.
Other analyses about the connection between classical random and scale-free networks have been reported in~\cite{krioukov01}. In the cited paper, the authors show that, for the cold regime, their network is scale-free, but as the temperature increases the network loses its metric structure and its hierarchical heterogeneous organization, becoming a classical random network.

Considering that many real complex networks are scale-free while others are not (see for example \cite{newman03}), we deem that the proposed fermionic model can be considered a good candidate for representing their evolution, at low and high temperatures.
As shown in Figure~\ref{fig:network_jump}, we analyzed also the dynamics of particles during both processes. In each simulation we observed that the cooling process takes more time to allow the particles to get to their final positions. During the cooling process, the number of particles changing position is very high from the first time step. In contrast, during the heating process, we found that, at the beginning, this number is small and rapidly increases after a few (usually about ten) time steps. Then, this number of jumps reaches a maximum and begins to decrease until all particles stop moving. We deem that this behavior is an effect of the Eq.~(\ref{eq:free_space_heat}), since it has been defined to prevent particles from occupying high-energy levels densely at high temperatures.

\section{Conclusions} \label{sec:conclusions}
IIn this paper, we have defined a fermionic network model that allows us to represent complex networks as quantum gases. Using this model, we have shown that network evolution is a temperature-dependent process characterized by three main phases: classical random, scale-free and \WTA.
On performing a cooling process, a transition from a classical random to a scale-free network takes place. Notably, the system achieves equilibrium when a \WTA~structure is reached, despite the non-equilibrium nature of the network evolution. On the other hand, on performing a heating process which starts from a \WTA~structure, the network evolution follows a slightly different path. In particular, a pure scale-free structure is not reached, although the actual structure is very similar.
Surprisingly, we found that the whole process, considering both cooling and heating, is not reversible when mapped to network evolution. Nevertheless, the separation between the classical random and \WTA~networks finds a full correspondence with the separation between classical and quantum regions for quantum gases. We deem that mapping complex networks to quantum systems can be useful for the analysis of some of their properties, including the possibility of modeling a competitive dynamics among nodes. As future work, we are planning to develop a model in which the fitness of each node can vary during the network evolution.

\end{document}